\documentclass[epj]{svjour}
\usepackage{graphics}
\usepackage{psfig}
\begin{document}
\title{\centerline{Energetic photoionization of neutral and ionic
metal clusters}}
\subtitle{~~~~~~~~~}
\author{\centerline{Himadri S Chakraborty \and M E Madjet}
}                     
\institute{\centerline{\small Max-Planck-Institut
f\"ur Physik Komplexer Systeme,
N\"othnitzer Strasse 38, D-01187 Dresden, Germany}}
\abstract{\small
We show, with an example of Na$_{92}$, that for 
jellium metal clusters the interference of 
fast electron-waves emitted from equivalent sites on the cluster edge produces
monochromatic oscillations  
in all photoionization observables as a function of the photoelectron
momentum;
the effect is equivalent to the usual dispersion phenomenon. 
In dealing with formalisms, a serious consequence of the  
inadequacy of self-interaction corrected
local density-functional theory in 
correctly accounting for the exchange interaction is identified. 
We also briefly
discuss the influence of the ionicity of the residual core  
on photospectra by considering the neutral member with $N=58$ and 
and the ionic member  with $N=52$ of the
Na$_{58}$ iso-jellium series, where $N$ is the number of
valence electrons.
A few final remarks on possible implications of these results on other 
quantum systems of delocalized electrons are made.\\ 
%
} 
\authorrunning{Himadri S Chakraborty and M E Madjet}
\titlerunning{Energetic photoionization of neutral and ionic metal clusters}
\maketitle
\section{Introduction}
\label{intro}
Due to the near-Coulomb character of a typical atomic potential with its strong
attractive behavior at close distances resulting electronic 
wavefunctions are generally compact around the nucleus. As a result, 
electrons in an atom are largely localized. There are two important
consequences of this property. 
First, since the gradient
of such a potential varies smoothly, electrons can receive the 
necessary recoil force ($-dV/dr$) to photoionize 
from `everywhere' within the spatial
extent of the system. Second, any rapid variation of the potential
within a small range
does not significantly affect photo-dynamical properties,
because the potential with its steep slope practically overwhelms
such local structures.     

On the other hand, atomic clusters 
exhibit properties that hover across the realm between the single
atom and the bulk. Contrary to the atomic context, where the
nucleus is practically dimensionless,  
a system of positive
core-ions with a certain spatial extension  
provides the necessary binding for  
valence electrons in a cluster. 
This background positive core can be approximated
as a homogeneously smeared out charge distribution, the jellium,
for a cluster of sufficiently large atoms with one or two valence
electrons per atom. Consequently,
the effective potential of such a system acquires, in comparison
to the atomic potential, a radically
different shape  
by having a flat interior and a sharp edge. We focus in this 
paper on this aspect of the cluster potential to scrutinize 
how does the shape influence the high energy photoionization behavior 
of a metal cluster and its ions. 
We further point out an unphysical consequence
of an {\it ad\,hoc} correction scheme 
that is popularly applied to the Kohn-Sham 
local-density-approximation (LDA) to account for 
spurious self-interactions. 

\section{Results and discussions}
\label{resdis}
Alkali-metal clusters having one valence electron per atom can be 
well described by a spherical jellium model which
entirely disregards the structure of the ionic core [1]. The Schr\"{o}dinger
equation can then be approximately solved by some ground state many
body theory considering Coulomb interactions among only 
valence electrons. As pointed out,
the resulting radial potential ($V$) typically behaves flat from
the origin upto almost the edge and then over a short range across
the edge ($r=R_{\mbox{c}}$) produces a steep barrier. 
This behavior physically implies that
the valence electron cloud is quasi-free 
over the most of the interior region
of the cluster but feels a rather strong confining
force in the vicinity of the surface. To understand the basic mechanism of the 
photoionization from such systems let us consider the dipole amplitude
of a typical $nl\rightarrow \epsilon l'$ transition. 
The radial matrix element
in the acceleration form of the
dipole interaction, that is proportional to the ionization recoil force,
is then $<\psi_{nl}|dV/dr|\psi_{\epsilon l'}>$; of course, the 
dual assumption of spherical symmetry and unpolarized photon is implied.
Now, the structure of $V$ clearly suggests
that its radial derivative must show a strong peak at the edge while is virtually 
vanishing elsewhere. This means that the predominant contribution to the 
overlap integral will come from the edge, which equivalently implies
that photoelectrons will largely emit from the surface region where it
receives enough recoil to eject. Strong contribution to the amplitude from
equivalent sites of the cluster edge will, therefore, show the signature 
of interference in the form of oscillations in the ensuing cross section
as well as in other physical parameters.   
While there has been no experimental study on metal clusters, oscillations
in the valence photoelectron intensity are indeed observed for C$_{60}$ [2] 
and C$_{70}$ [3] fullerine molecules, which are also 
systems of quasi-free de-localized 
electrons confined within a nearly spherical shell. 
%
\begin{figure*}
\resizebox{0.75\textwidth}{!}{%
\psfig{figure=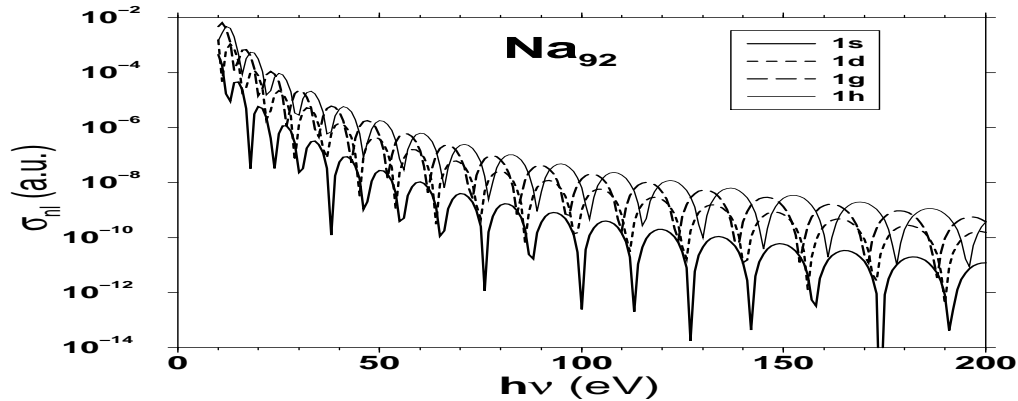,height=3.2in,width=8in}
}
\caption{1s, 1g, and 1h subshell photoionization  
cross sections as a function of photon-energy calculated in LDA-SIC.}
\label{fig:1}       
\end{figure*}

Situation begins to get simpler as the energy of the incident photon
increases. In general, the continuum radial wavefunction $\psi_{\epsilon l'}$
for such potentials are very nearly the spherical Bessel function $j_{l'}(k_{nl}r)$,
where $k_{nl}$ is the photoelectron momentum.
With high photon-energy, and therefore 
fast photoelectrons, this function limits to $\cos (k_{nl}r+l'\pi/2)$,
the usual first Born picture. Insertion of this high energy form into the
radial matrix element immediately reveals that the oscillation
in the dipole amplitude will have a frequency close to the radius of the cluster.
Evidently, the cross section, which is the sum of 
the squared modulous of amplitudes of allowed
transitions, will then oscillate in a frequency approximately the diameter:
\begin{equation}
\sigma_{nl} \sim \cos(2R_{\mbox{c}}k_{nl}+l'\pi)
\end{equation}
In fact, the oscillation in alkali-metal clusters has been predicted earlier
through a kind of semi-classical study [4] and has later been verified by 
numerical calculations of the total cross section of Na$_{40}$
using some effective mean-field LDA potential [5].
%
\begin{figure*}
\resizebox{0.75\textwidth}{!}{%
\psfig{figure=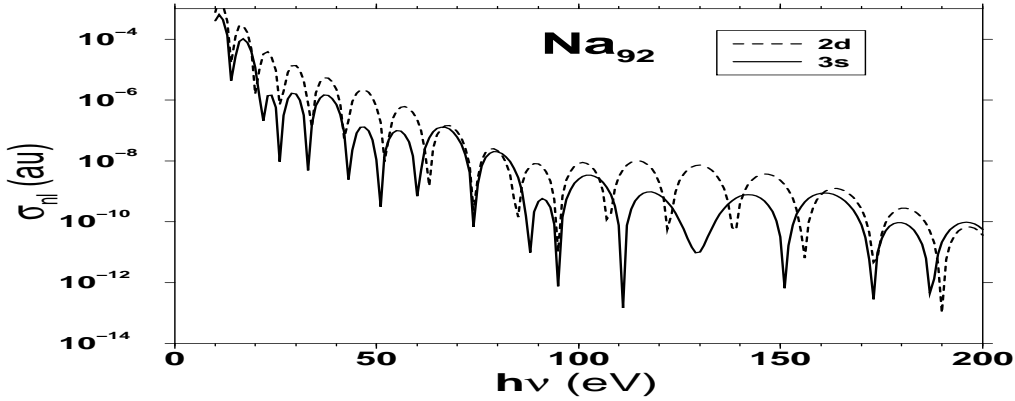,height=3.2in,width=8in}
}
\caption{Same as Figure 1 but for 2d and 3s electrons
}
\label{fig:2}       
\end{figure*}

In this study, we calculate subshell photoionization cross sections
for Na$_{92}$ cluster 
(1s$^2$1p$^6$1d$^{10}$2s$^2$1f$^{14}$2p$^6$1g$^{18}$2d$^{10}$ 1h$^{22}$3s$^2$)
from threshold to about 1 KeV photon-energy in an independent
particle frame. The ground
state is described by the electron self-interaction corrected
local-density approximation (LDASIC) [6], that approximately restores
the physical long range $-1/r$ behavior of the resulting single-electron
potential. Figure 1 presents some selected subshell cross sections
from $n=1$ manifold. The oscillation in cross sections is very much in
evidence and the Fourier transform (not shown) 
of the cross sections (as a function of the respective $k_{nl}$) 
indeed peaks at a position which is approximately
the diameter of the cluster.    

An interesting physical analogy of this phenomenon  
can apparently be drawn
with the usual diffraction mechanism in optics. The light shines on a 
metal cluster which is like a spherical double-slit with slit width to be the
cluster diameter. `Free' delocalized electrons over the bulk of the interior
region are nearly insensitive to the light. Photoelectron 
waves emanate only from the two slits, which effectively are the 
diametrically opposite sites on the surface of the sphere. These secondary waves
with a certain path difference, that is equal to the slit width and hence
the diameter, subsequently interfere. As a result,
the cross section oscillates in a frequency $2R_{\mbox{c}}$, a feature
equivalent to  
the distance between consecutive fringes in a diffraction pattern 
being $2\pi(2R_{\mbox{c}})$.

For a ground state subshell of the angular momentum $l$ the two dipole-allowed
free angular momenta are $l'=l\pm1$. Therefore, it can be easily 
verified by using equation (1) that
the phase contribution to any $\sigma_{nl}$ from the continuum wavefunctions
is always $2\pi$ at high enough energy. This implies that the 
high energy phase difference between any two subshell cross sections will be 
predominantly determined by the phase contribution from their ground state angular
momentum $l$. Thus, for any two subshells with the difference between their $l$ values being 
an odd integer the corresponding cross sections will show oscillations 
largely out of phase, while the difference being even they will oscillate
in phase. This behavior is clearly evident among the cross sections
in Figure 1; while a rough phase agreement is seen between 1s and 1d cross sections, 
1g and 1h are clearly out of phase. 
%
\begin{figure}
\resizebox{0.75\textwidth}{!}{%
\psfig{figure=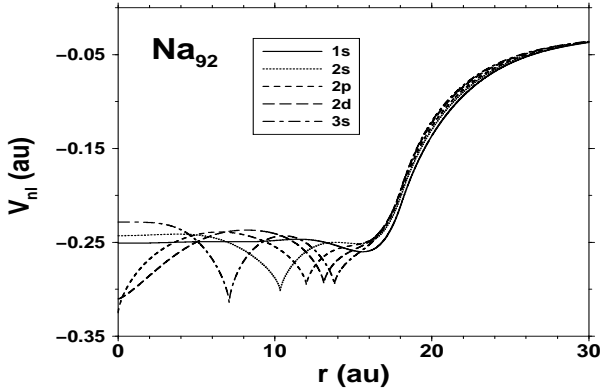,height=0.04in,width=0.06in}
}
\caption{LDASIC state dependent radial potentials for some selected
subshells}
\label{fig:3}       
\end{figure}

Figure 2 shows the cross sections for 2d and 3s subshells. 
Surprisingly, oscillations in these cross sections seem to have more 
than one frequency. Indeed, the corresponding Fourier spectra 
identify several other frequencies besides the physical 
one that connects to the
cluster diameter. These extra frequencies can be shown to be entirely
spurious and originate from the unphysical cusp that the self-interaction 
correction (SIC) generates in subshell-dependent radial potentials
at the position where the respective single 
electron density attains nodal zero [7]. 
Figure 3 shows some of these potentials. 
Subshells with $n=1$ are nodeless and, therefore,
the 1s potential is seen to have the usual shape. 
This fact explains the correct
behavior of the resulting cross section for $n=1$ family.   
Subshells with
$n=2$, containing one node (Figure 3), 
yields a second frequency in the dipole amplitude
that interferes with the physical frequency to contaminate the cross section
with three spurious frequencies. 
The same mechanism works for $n=3$ subshell which has
two nodal points (Figure 3) to eventually 
produce eight unphysical frequencies 
in the photo cross section. We found that this is an artifact of an
external, and hence approximate,
implementation of SIC in a local frame [7]. The correct
elimination of the self-interaction must treat {\em fully} the exchange effect
which is intrinsically non-local. This answers why SIC is automatically 
built in the non-local Hartree-Fock formalism. 
The LDASIC is certainly a powerful
tool to handle the low energy dynamics [6], but it is also true 
that the effect discussed here 
remains virtually recessive in this energy range for this
method, because the continuum wave
with the longer wavelength can hardly `detect' this 
structure in the potential. On the other hand, at higher energies this intrinsic
limitation of the formalism causes major problem, although
SIC is found largely unimportant over the high energy range
since with the contribution
to the overlap integral predominantly coming from the edge 
any refinement in the long distance character of 
the wavefunction leaves the result practically unaltered. 
It is, therefore, of crucial importance
to choose appropriate theoretical techniques to avoid mis-interpretation of
photo-dynamical results for metal clusters.  
%
\begin{figure}
\resizebox{0.75\textwidth}{!}{%
\psfig{figure=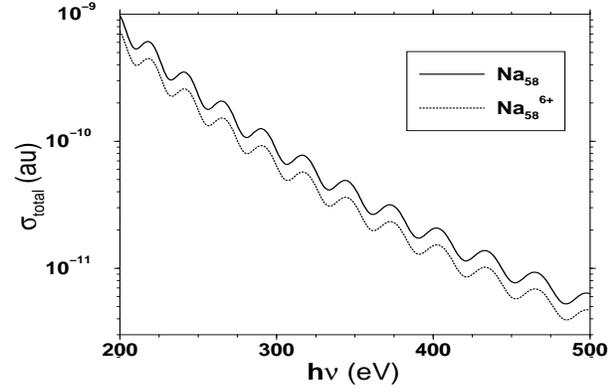,height=0.04in,width=0.06in}
}
\caption{Total high-energy cross sections for Na$_{58}$ and 
Na$_{58}^{6+}$ systems calculated in usual LDA without SIC
incorporated
}
\label{fig:4}       
\end{figure}

It is often of interest to examine
the influence of the initial ionization state on the 
photoionization dynamics. A possible approach to this direction 
is to look at the change in the behavior of photo cross sections
as a function of the residual charge by reducing the
valence electron number ($N$) but keeping 
the charge of the jellium background 
constant (iso-jellium series). 
In the atomic regime, recent investigations on isoelectronic
sequences revealed a number of important effects at low photon-energies [8].
We consider here the photoionization of neutral Na$_{58}(N=58)$ and  
Na$_{58}^{6+}(N=52)$ ion to make a preliminary study on the effect of 
the residual core ionicity on the total ionization cross section; a 
detailed study including subshell cross sections
for several other ions along both iso-jellium and isoelectronic
series will be 
considered elsewhere. 

Total cross sections
for both the systems calculated via LDA, but without SIC 
incorporated, are compared in Figure 4.
Both the cross sections look qualitatively similar having the 
identical trend in terms of oscillations. 
Of course, with the increase of positive residual
ionicity bound electrons find it harder to escape, which explains                    
why the ionic cross section is lower.
At this point it may be rather interesting to look
at the LDA ground state radial 
potentials for both the systems, as are plotted 
in Figure 5. Clearly, the increase of the residual positive charge
from $+1$ in the neutral to $+7$ in the ion only
deepens the potential
while the position and the softness of the edge  
remains largely
unchanged. Although this is a direct consequence of
the jellium approximation, but a proper inclusion of
the ionic dynamics is not expected to change the situation 
significantly. It is evident that the
identical frequency of the oscillation in both the cross sections
is owing to the unaltered position of the cluster edge. Also, the
identical softening at the edge of the potentials accounts for the similar
background decay pattern of the cross sections.   
What is rather interesting, however, is while the depth of the potential 
for the ion increases greatly, by about 200\% of that of its neutral
partner, 
respective total cross sections (Figure 4) 
are very close to each other! In fact, over the entire
high energy range they differ only by a 
simple energy-independent scaling factor that is  
 roughly close to unity. 

It turns out that this scaling quantity is a function of $z$, the
initial ionization of the system.
Starting from the fact that the dominant contribution to the cross section
comes from the cluster edge, where the derivative of the potential
peaks, it immediately fallows that the $z$-dependence of the cross section
will be predominantly determined by the $z$-dependence of the bound state 
wavefunction at the edge. Therefore, cross sections must be of 
similar strength as the magnitude of radial waves are
considerably close to each other at the edge, no matter
what positive charge they experience. Thus, the situation clearly
offers a  possibility of 
generating the theoretical ionic photoionization data over the high energy range
through a simple scaling of the result for respective neutral member;
the formulation of a detailed scheme is on the anvil.   
%
\begin{figure}
\resizebox{0.75\textwidth}{!}{%
\psfig{figure=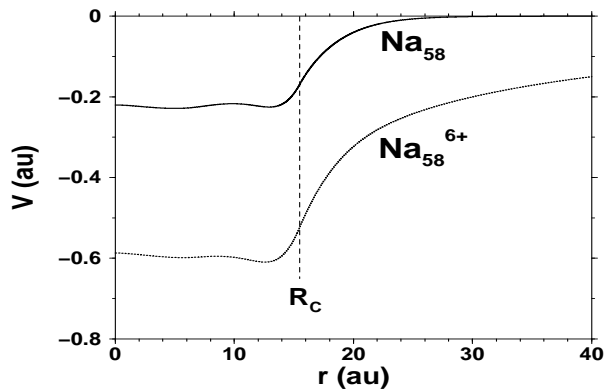,height=0.04in,width=0.06in}
}
\caption{LDA radial potentials for the systems as in Figure 4}
\label{fig:5}       
\end{figure}
\section{Concluding remarks}
\label{con}
Energetic photo-spectra of metal clusters are largely dominated 
by oscillations from a diffraction-like mechanism through
the interference
of electron waves emitted by the cluster edge. This behavior
is directly connected to the delocalized character of the electron
charge distribution. Similar mechanism
is also found operative for the valence photoionization
of C$_{60}$ molecule [9]. A unique
structural feature of C$_{60}$ is a hollow cage that can easily
hold an atom inside; in fact such
endohedral confinement has been made possible
in the laboratory. Therefore, having understood
the mechanism of oscillations in C$_{60}$ photo-spectra 
it is now of considerable interest to study the influence of C$_{60}$ cavity
on the 
photo-response of an atom endohedrally confined into it and the
associated spectroscopy. Investigations along this line is
underway and will be published soon [10]. Furthermore, with the enormous
advent of semiconductor technology it has been possible to prepare
small electron-islands of
strong quantum properties, quantum dots, within the semiconductor 
lattice. Understanding of the physics of quantum dots is valuable 
because of the potentially huge applicational importance of these
objects. Electrons
forming a quantum dot are certainly delocalized in nature and, therefore,
their ionization by photon impact with high enough energy may
show diffraction pattern providing valuable information about
the hither-to-unknown geometry of the dot. A part of our current
endeavor is focussed on this subject. 
\\
\\
We thank Professor Jan-M Rost for many useful discussions.
%
%

\end{document}